\documentclass[11pt,a4paper]{article}
\usepackage{jcappub}
\usepackage{booktabs}
\usepackage{bm}
\usepackage{graphicx}
\graphicspath{{figures}}

\newcommand{\lcdm}{\Lambda\text{CDM}}
\newcommand{\kmsmpc}{\ensuremath{\mathrm{km\,s^{-1}\,Mpc^{-1}}}}
\newcommand{\Om}{\Omega_{\rm m}}
\newcommand{\Ode}{\Omega_{\Lambda}}
\newcommand{\Ok}{\Omega_{k}}

\newcommand{\Ogamma}{\Omega_{\gamma}}
\newcommand{\Hzero}{H_0}

\newcommand{\eff}{\mathrm{eff}}

\title{Hierarchical Gaussian-process reconstruction of late-time expansion with DESI BAO and compressed Planck priors}

\author{Enkun Li\note{Corresponding author.},}
\author{Yu-Hao Mu,}
\author{and Zhihuan Zhou}
\affiliation{School of Engineering, Xi'an International University,\\
Xi'an 710077, People's Republic of China}
\emailAdd{lekf123@163.com}

\abstract{We reconstruct the late-time expansion history with a hierarchical Gaussian process (GP) that co-samples the kernel hyperparameters $(\sigma_f,l)$ together with $(\Hzero,\Om,\Ok,\omega_b h^2)$. The data are 37 cosmic-chronometer $H(z)$ points, DESI DR2 BAO, Planck 2018 compressed distance priors, and Pantheon+ supernovae, coupled by a Monte-Carlo effective likelihood. DESI's headline Chevallier--Polarski--Linder (CPL) preference at $2.8$--$4.2\sigma$ uses the full Planck likelihood; we do not reproduce or refute that result. On this compressed-CMB pipeline the baseline posterior is $w(z\simeq 0)=-0.80^{+0.26}_{-0.23}$ (68\%~C.L.), less than $1\sigma$ from $w=-1$, with $l=3.79^{+0.81}_{-1.12}$. A CPL fit on the same data improves nested $\lcdm$ by only $\Delta\chi^2\sim 1$; Akaike's criterion prefers nested $\lcdm$ ($\Delta\mathrm{AIC}\simeq 3$), while BIC is quoted only as an indicative check for correlated supernovae. The reconstructed $f_{\rm DE}=1$ lies inside the 68\% band at every grid point with $z\le 1.6$. Ablations that fix $(\sigma_f,l)$, drop SN or LRG1/2 BAO, replace the radial-basis kernel by a Mat\'ern family, swap the SN catalogue (Union3, DES-Y5, DES-Dovekie), or nest a residual GP around $\lcdm$ leave the median $w(0)$ within $\lesssim 1\sigma$ of $-1$.}

\arxivnumber{2607.23593}

\begin{document}
\maketitle
\section{Introduction}
% =============================================================================

Late-time cosmic acceleration is commonly attributed to a cosmological
constant with $w=-1$
\cite{Riess:1998cb,Perlmutter:1998np,Schmidt:1998ys,Weinberg:1988cp,Copeland:2006wr}.
Observational analyses usually start from the Chevallier--Polarski--Linder
(CPL) form
\begin{equation}
w(a)=w_0+w_a(1-a)
\label{eq:cpl}
\end{equation}
\cite{Chevallier:2000qy,Linder:2002et}. Any fixed functional shape can bias
inferred $w(z)$ if the true expansion history lies outside that family.

Non-parametric reconstructions avoid prescribing $w(z)$ a priori.
Gaussian-process (GP) regression
\cite{Holsclaw:2010sk,Seikel:2012uu,Shafieloo:2012ht,Rasmussen:2006}
is the tool used here; principal-component reconstructions
\cite{Huterer:2002hy,Albrecht:2006um,Crittenden:2005wj} are complementary.

The Dark Energy Spectroscopic Instrument (DESI) Data Release~2 provides
13~baryon acoustic oscillation (BAO) distance measurements from seven tracers
over $0.1<z<4.2$ \cite{DESI:2025,DESI:2025zgx,DESI:2025zpo}, building on DR1
\cite{DESI:2024mwx,DESI:2024uvr}. In the Chevallier--Polarski--Linder (CPL)
parameterisation, DESI reports a preference for a quintessence-to-phantom
transition when BAO are combined with CMB and supernovae, at a significance
that depends on the CMB likelihood and the external data combination
\cite{DESI:2025}. Follow-up reconstructions and critical assessments include
\cite{Lodha:2025qbg,Gu:2025xie,Zhang:2025dwu,Cortes:2024lgw,Wolf:2023uno,%
Efstathiou:2024xcq,Calderon:2024uwn,Lodha:2024kty,Li:2025owk,Wang:2025bkk}.
Analyses that replace the full Planck likelihood by compressed distance
priors, or that combine chronometers with BAO, often find a weaker dynamical
preference
\cite{Loubser:2025fxr,Cortes:2025joz,Wang:2025vfb,Ye:2025ark,deSouza:2025rhi}.
The quoted significance is also sensitive to the assumed $w(z)$
parameterisation \cite{Zhang:2025,Li:2025}.
Gaussian-process reconstructions on similar late-time stacks with compressed
CMB information typically recover $w(z)\simeq-1$ within uncertainties, with at
most a supernova-dependent $1\sigma$ crossing hint
\cite{Maqsood:2026gpd}, while a density-space GP analysis finds that
$\lcdm$, thawing quintessence, and CPL all remain compatible with the
reconstructed $\rho_{\rm DE}(z)$ at 95\% confidence
\cite{deSouza:2025rhi}. Joint expansion--growth GPs likewise report no
significant dynamical dark energy or dark-sector interaction
\cite{You:2026dsd}. Kernel smoothness itself acts as a prior that can
systematically shift reconstructed $\Hzero$ along the Mat\'ern
differentiability hierarchy \cite{Maqsood:2026ker}. Calibration-independent
BAO--SN tests further show that the supernova sample is a first-order
systematic \cite{Dinda:2025jcap,Dinda:2026upd}; reconstructions that vary
both the kernel and the SN catalogue can move a $w\simeq-1$ result to a
$2$--$4\sigma$ preference \cite{Maqsood:2026thermo}.
Our aim is not to reproduce or refute DESI's full-likelihood CPL result.
We ask how a hierarchical GP, and a set of probe and kernel choices, move
$w(z)$ and $f_{\rm DE}(z)$ when the CMB information is restricted to compressed
distance priors.

Reconstructing $w(z)$ from data typically means inferring $H(z)$ or $d_L(z)$
non-parametrically and then differentiating. GP regression supplies a
posterior over the target function and its derivatives
\cite{Rasmussen:2006,Seikel:2012,Benisty:2023}. In practice, most GP
dark-energy studies fix the kernel hyperparameters $(\sigma_f,l)$ at their
maximum-likelihood (ML) values before propagating $H(z)$ uncertainties
\cite{Seikel:2012,Holanda:2017,Zhang:2018gjb}. That two-step procedure omits
hyperparameter uncertainty in the final error budget
\cite{Crittenden:2011aa,Zhao:2017cud,Wang:2018fng}. Cosmic chronometers (CC),
BAO, CMB, and Type~Ia supernovae (SN~Ia) further enter through different
non-linear functionals of $H(z)$
\cite{Jimenez:2001gg,Eisenstein:2005su,Planck:2018vyg,Brout:2022vxf}; many
pipelines combine them with a fixed-background $\chi^2$ rather than
co-sampling $(\sigma_f,l)$ together with $\theta$
\cite{Zhang:2025,Li:2025}.
The architecture used here is still not a fully joint latent-$H$ model on
all probes: the GP covariance is conditioned on chronometers alone
(Sec.~\ref{sec:method}).

Here we apply a hierarchical GP to the multi-probe combination
CC+DESI~DR2+Planck compressed distance priors+Pantheon+, and we test how the
inferred $w(z)$ responds to concrete analysis choices: (i)~co-sampled versus
fixed $(\sigma_f,l)$; (ii)~the SN catalogue (Pantheon+, Union3, DES-Y5,
DES-Dovekie, or none); (iii)~removal of the LRG1/2 BAO bins;
(iv)~the Mat\'ern differentiability sequence $3/2$--$9/2$ versus RBF, and a
tightened prior on the length scale $l$; and (v)~a residual GP nested on
$\lcdm$. The deliverable is that response under compressed CMB information.
The Hubble constant $\Hzero$ is left free (no SH0ES or Planck $H_0$ prior), so
the posterior may sit between the CMB and local distance-ladder values
\cite{Planck:2018vyg,Riess:2021jrx,Vagnozzi:2023nrq} without being forced
either way. The technical ingredients are co-sampled hyperparameters in a
Markov-chain Monte Carlo (MCMC) jointly with
$(\Hzero,\Om,\Ok,\omega_b h^2)$ \cite{ForemanMackey:2013,Goodman:2010}, an
effective likelihood $\mathcal{L}_{\eff}$ obtained by averaging BAO+CMB+SN
over GP draws of $H(z)$, a same-data CPL comparison under the same compressed
Planck priors, and the normalised dark-energy density
$f_{\rm DE}(z)=\rho_{\rm DE}(z)/\rho_{\rm DE}(0)$.

Section~\ref{sec:theory} fixes notation. Section~\ref{sec:method} describes
the data, hierarchical model, and sampler. Section~\ref{sec:results} reports
parameter constraints, reconstructions, and framework tests.
Section~\ref{sec:disc} discusses the results; Sec.~\ref{sec:conc} concludes.

% =============================================================================
\section{Theoretical framework}
\label{sec:theory}
% =============================================================================

\subsection{Background cosmology}

We work in a flat or weakly curved Friedmann--Lema\^{i}tre--Robertson--Walker
background. The Hubble rate satisfies
\begin{equation}
H^2(z) = \Hzero^2\big[\Ogamma(1+z)^4 + \Om(1+z)^3 + \Ok(1+z)^2
+ \Omega_{\rm DE}(z)\big],
\label{eq:friedmann2}
\end{equation}
with radiation density $\Ogamma=2.47\times10^{-5}h^{-2}$ and
$h=\Hzero/(100\,\kmsmpc)$. In $\lcdm$,
$\Omega_{\rm DE}=1-\Om-\Ok-\Ogamma$ is constant and $w=-1$.

Distances are obtained from the line-of-sight comoving coordinate
\begin{equation}
r(z) = \int_0^z \frac{c\,dz'}{H(z')}.
\end{equation}
The transverse comoving distance $D_M(z)$, Hubble distance
$D_H(z)=c/H(z)$, luminosity distance $d_L(z)=(1+z)\,r(z)$, and angular-diameter
distance $D_A(z)=r(z)/(1+z)$ follow in the usual way (with the appropriate
curvature functions when $\Ok\neq 0$). BAO observables are ratios of these
distances to the drag-epoch sound horizon $r_d$.

Dark energy is characterised by
\begin{equation}
w(z) \equiv \frac{p_{\rm DE}(z)}{\rho_{\rm DE}(z)}.
\end{equation}
Energy conservation then implies
\begin{equation}
\Omega_{\rm DE}(z) = \Omega_{\rm DE,0}\,
\exp\!\Big(-3\!\int_0^z \frac{1+w(z')}{1+z'}\,dz'\Big).
\label{eq:OmegaDE}
\end{equation}
Given a reconstructed $H(z)$, it is often more stable to work with the
normalised density
$f_{\rm DE}(z)\equiv\Omega_{\rm DE}(z)/\Omega_{\rm DE,0}$ obtained by
inverting Eq.~\eqref{eq:friedmann2}, and then to form
\begin{equation}
w(z) = -1 + \frac{1+z}{3}\,\frac{f'_{\rm DE}(z)}{f_{\rm DE}(z)},
\label{eq:wz2}
\end{equation}
where a prime denotes $d/dz$. The ratio $f'/f$ amplifies derivative noise,
so $f_{\rm DE}(z)$ (equivalently $\rho_{\rm DE}$) is the more directly
constrained background diagnostic
\cite{deSouza:2025rhi}. Equation~\eqref{eq:wz2} is the relation used in
the reconstruction pipeline below.

\subsection{Model-independent reconstruction}

CPL and related parameterisations condition the inference on a fixed $w(z)$
family [Eq.~\eqref{eq:cpl}]. Here $H(z)$ is treated as a latent function drawn
from a GP \cite{Rasmussen:2006,Sarkka:2019book}, and $w(z)$ follows from
Eq.~\eqref{eq:wz2}. The GP posterior also supplies the derivatives needed for
that conversion \cite{Seikel:2012,Holanda:2017,Benisty:2023}. The amplitude
$\sigma_f$ and correlation length $l$ of the kernel are treated as nuisance
parameters to be marginalised
\cite{Crittenden:2011aa,Zhao:2017cud,Wang:2018fng}.

The covariance kernel is a smoothness prior, not a physical model of dark
energy. Along the Mat\'ern family, increasing mean-square differentiability
systematically lowers reconstructed $\Hzero$ and suppresses local slope
features \cite{Maqsood:2026ker}. The squared-exponential (RBF) kernel is the
infinitely differentiable endpoint of that hierarchy. The hierarchical
pipeline reports the sequence $\nu=3/2,5/2,7/2,9/2$ and RBF as a smoothness
prior, not as a test of which kernel is ``true''.

Production reconstructions use a \emph{zero-mean} GP on $H(z)$ itself,
\begin{equation}
H(z)\sim\mathcal{GP}\bigl(0,k(z,z';\sigma_f,l)\bigr).
\label{eq:zeromean}
\end{equation}
That is a modelling choice with physical content, not a neutral statistical
default. The baseline posterior amplitude is
$\sigma_f=284.6^{+135.9}_{-72.2}\,\kmsmpc$ (Sec.~\ref{sec:results}), a large
fractional perturbation of $H(z)$ at low redshift. Friedmann inversion of
each draw then \emph{defines} a fluid, and $w(z)$ follows from $f'_{\rm DE}/f_{\rm DE}$.
Alternatives include a GP on $\ln H$, a GP on $f_{\rm DE}-1$, or a residual
process around a cosmological-constant background,
\begin{equation}
H(z)=H_\Lambda(z\mid\theta)+f(z),\qquad
f\sim\mathcal{GP}(0,k),
\label{eq:residual}
\end{equation}
so that $\sigma_f\to 0$ recovers $\lcdm$ at fixed $\theta$.
Equation~\eqref{eq:residual} is implemented as a nested check
(ablation-length MCMC) and is not used to replace the production numbers.
The residual chain pulls $w(0)$ back toward $-1$ and drives $\Hzero$ onto
the lower prior edge (Sec.~\ref{sec:framework}), which shows that part of
the baseline offset from $\Lambda$ is prior-supported rather than
data-forced. We do not quote a Bayes factor between
Eq.~\eqref{eq:residual} and Eq.~\eqref{eq:zeromean}.

Because BAO, CMB, and SN~Ia depend non-linearly on $H(z)$, a point estimate of
the GP mean is not sufficient for a joint likelihood. The hierarchical model of
Sec.~\ref{sec:method} therefore marginalises the latent expansion history by
Monte-Carlo integration while sampling the GP hyperparameters and the
cosmological parameters together.

% =============================================================================
\section{Method}
\label{sec:method}
% =============================================================================

\subsection{Data sets}

\textit{Cosmic chronometers.}
We use 37 differential-age $H(z)$ measurements compiled in
Ref.~\cite{Moresco:2022}, spanning $0.07<z<1.965$. These points provide the
only direct anchors of $H(z)$ in the analysis: the GP covariance is
conditioned on them alone. BAO, CMB, and SN update the cosmological and
kernel parameters through the effective likelihood below; they do not
re-condition the GP on those distance data. Consequently the small-scale
structure of $H(z)$ is set by 37 chronometer points, not by a joint latent
process on BAO or SN distances. Measurement uncertainties are
taken as published and treated as diagonal in the GP noise covariance.

\textit{DESI DR2 BAO.}
Thirteen BAO measurements from BGS, LRG, ELG, and QSO tracers cover
$0.1\lesssim z\lesssim 4.2$ \cite{DESI:2025}. Depending on the tracer, the
data vector consists of $(D_M/r_d,\,D_H/r_d)$ with the published correlation
coefficient, or of the isotropic combination $D_V/r_d$. We employ the full DR2
covariance among the 13 rows. In one ablation we remove the LRG1 and LRG2 bins
to test sensitivity to the intermediate-redshift BAO leverage emphasised in
some DESI discussions.

\textit{Planck 2018 CMB distance priors.}
Instead of the full Planck temperature and polarisation likelihood, we use
the compressed distance priors $(R,\ell_A,\omega_b h^2)$ and their
$3\times 3$ covariance derived from TT,TE,EE+lowE
\cite{Huang:2020,Planck:2020}, where
\begin{equation}
R=\sqrt{\Om\Hzero^2}\,D_M(z_*)/c,\qquad
\ell_A=\pi D_M(z_*)/r_s(z_*)
\end{equation}
at the photon-decoupling redshift $z_*\simeq 1090$. This choice keeps the
pipeline computationally light and matches several recent chronometer+BAO
studies, but it is \emph{not} equivalent to DESI's baseline combination with
the full Planck likelihood. All scientific claims below are therefore
restricted to this compressed-CMB setting
\cite{Cortes:2025joz,Wang:2025vfb,Ye:2025ark}.

\textit{Pantheon+.}
The Hubble-flow Pantheon+ sample comprises 1624 SN~Ia
\cite{Brout:2022vxf} with $0.01<z<2.3$. We use the full
$1624\times 1624$ statistical-plus-systematic covariance matrix. The absolute
magnitude is marginalised analytically following the standard quadratic
completion of the distance-modulus $\chi^2$
\cite{Goliath:2001,Conley:2011}.
This Hubble-flow selection ($N=1624$) differs from the
$z_{\mathrm{HD}}>0.01$ cut ($N=1590$) used in some compressed-CMB CPL
analyses of the same DESI~DR2 tables; SN vectors should not be treated as
identical across those pipelines.

\textit{Union3, DES-Y5, and DES-Dovekie.}
To test SN calibration as a first-order systematic
\cite{Dinda:2025jcap,Dinda:2026upd,Maqsood:2026thermo}, we also run the same
compressed pipeline with Union3 (22 Hubble bins)
\cite{Rubin:2023ovl}, the DES 5-year Hubble diagram
\cite{DES:2024jxu}, and the recalibrated DES-Dovekie sample
\cite{Popovic:2025dovekie}. Absolute magnitude is marginalised in the same way.

\subsection{Hierarchical model architecture}

The model has three layers:
\begin{enumerate}
\item \textbf{Parameters:}
$\theta = \{ \Hzero, \Om, \Ok, \omega_b h^2, \ln\sigma_f, \ln l \}$.
\item \textbf{Latent process:}
$H(z)$ is a zero-mean GP with kernel $k(z,z')$, conditioned on the CC data.
\item \textbf{Observables:}
BAO, CMB, and SN~Ia are non-linear functionals of $H(z)$ and $\theta$.
\end{enumerate}

The joint posterior is
\begin{equation}
\begin{aligned}
&p(\theta\,|\,\mathrm{CC},\mathrm{BAO},\mathrm{CMB},\mathrm{SN}) \propto  \\
& ~~~~ p(\mathrm{CC}\,|\,\sigma_f,l) \times
\mathcal{L}_{\eff}(\mathrm{BAO},\mathrm{CMB},\mathrm{SN}\,|\,\theta)
 \times \pi(\theta),
\label{eq:joint_posterior}
\end{aligned}
\end{equation}
where $p(\mathrm{CC}\,|\,\sigma_f,l)$ is the GP log-marginal likelihood
(analytically tractable) and $\mathcal{L}_{\eff}$ is the Monte-Carlo effective
probe likelihood defined below. Flat priors $\pi(\theta)$ are listed in
Table~\ref{tab:priors}.

\subsection{GP conditioning on cosmic-chronometer data}

For a zero-mean GP with kernel $k(z,z')$ and hyperparameters $(\sigma_f,l)$,
the posterior mean and covariance at test redshifts $\mathbf{z}_*$, given
$N_{\rm CC}$ data points $(\mathbf{z},\mathbf{H},\boldsymbol{\sigma}_H)$, are
\begin{align}
\boldsymbol{\mu}_* &= \mathbf{K}_{*N}(\mathbf{K}_{NN}+\boldsymbol{\Sigma})^{-1}\mathbf{H},
\\
\boldsymbol{\Sigma}_* &= \mathbf{K}_{**}-\mathbf{K}_{*N}(\mathbf{K}_{NN}+\boldsymbol{\Sigma})^{-1}\mathbf{K}_{N*},
\end{align}
where $\mathbf{K}_{NN}$ is the $N_{\rm CC}\times N_{\rm CC}$ kernel matrix,
$\mathbf{K}_{*N}$ is the $M\times N_{\rm CC}$ cross-covariance,
$\boldsymbol{\Sigma}=\mathrm{diag}(\sigma_H^2)$, and $M$ is the number of test
points. The log-marginal likelihood is
\begin{equation}
\ln p(\mathrm{CC}\,|\,\sigma_f,l) =
-\tfrac12\mathbf{H}^{\!\top}\boldsymbol{\alpha}
-\sum_i\ln L_{ii}
-\tfrac12 N_{\rm CC}\ln(2\pi),
\label{eq:log_ml}
\end{equation}
where $\mathbf{L}\mathbf{L}^{\!\top}=\mathbf{K}_{NN}+\boldsymbol{\Sigma}$ is
the Cholesky factor and $\boldsymbol{\alpha}$ solves
$\mathbf{L}\mathbf{L}^{\!\top}\boldsymbol{\alpha}=\mathbf{H}$. A small jitter
($\sim 10^{-10}$ on the diagonal, increased to $10^{-6}$ if needed) stabilises
the factorisation.

The baseline kernel is the squared-exponential (RBF) form
\begin{equation}
k(z,z')=\sigma_f^2\exp\!\big[-(z-z')^2/(2l^2)\big].
\end{equation}
Mat\'ern kernels and related alternatives are discussed in
\cite{Rasmussen:2006,Zhang:2018gjb,GomezValent:2018hwc,Maqsood:2026ker}; the
sequence $\nu=3/2$--$9/2$ is reported in Sec.~\ref{sec:framework} as a
smoothness-prior test. Predictions are evaluated on a uniform grid $z_*\in[0,2.5]$ with $M=200$
points. Beyond $z=2.5$ the
normalised dark-energy density is held fixed when distances to recombination
are required, so that the late-time GP does not invent unconstrained
high-redshift structure.
For a non-GP reader: high-$z$ $w(z)$ is therefore \emph{not} reconstructed
from the CMB. Early-time physics enters only through $r_s$ and the compressed
distance priors.

Optionally the GP may be conditioned on residuals
$\mathbf{H}-\mathbf{H}_\Lambda(\theta)$ as in Eq.~\eqref{eq:residual}. In the
$\sigma_f\to 0$ limit the residual mean vanishes (verified at the
chronometer layer to $\lvert\langle f\rangle\rvert\lesssim 10^{-6}\,\kmsmpc$).
Production baseline chains use the zero-mean GP on $H(z)$. A residual chain
with prior $\sigma_f\in[0.1,500]\,\kmsmpc$ (so that nested $\lcdm$ sits at
the lower edge) is reported in Sec.~\ref{sec:framework}.

\subsection{Effective likelihood via Monte-Carlo integration}

BAO predictions require $D_M(z)$ and $D_H(z)$ (or $D_V$), CMB distance priors
require $D_M(z_*)$ and $r_s(z_*)$, and SN~Ia require $d_L(z)$. Because $H(z)$
is latent, these probes are coupled through the GP posterior. We define
\begin{equation}
\mathcal{L}_{\eff}(\theta) =
\mathbb{E}_{H|\mathrm{CC},\theta}\!
\big[\, \mathcal{L}_{\rm BAO}(\theta,H)\, \mathcal{L}_{\rm CMB}(\theta,H)\,
\mathcal{L}_{\rm SN}(\theta,H)\,\big].
\label{eq:Leff}
\end{equation}
The expectation is estimated with $N_s$ independent draws
$\{H^{(i)}(z_*)\}_{i=1}^{N_s}$ from the GP posterior (via the
eigen-decomposition of $\boldsymbol{\Sigma}_*$),
\begin{equation}
\ln \mathcal{L}_{\eff}(\theta) \approx
\ln\!\sum_{i=1}^{N_s} e^{-\chi^2_i/2} - \ln N_s,
\label{eq:Leff_mc}
\end{equation}
where
$\chi^2_i = \chi^2_{\rm BAO}(\theta,H^{(i)}) + \chi^2_{\rm CMB}(\theta,H^{(i)})
+ \chi^2_{\rm SN}(\theta,H^{(i)})$ and each $\chi^2$ uses the corresponding
data covariance. Production chains use $N_s=20$--$30$ draws per likelihood
evaluation; this Monte-Carlo average is the dominant computational cost once
the Pantheon+ covariance is included. Equation~\eqref{eq:Leff} updates
$\theta$, not the GP kernel matrix: each draw is still the CC-conditioned
posterior at the current $(\sigma_f,l)$.

For each draw, $f_{\rm DE}(z)$ is obtained by inverting
\begin{equation}
E^2(z) = \Ogamma(1+z)^4 + \Om(1+z)^3 + \Ok(1+z)^2 + \Ode\,f_{\rm DE}(z),
\label{eq:friedmann}
\end{equation}
with $E=H/\Hzero$ and $\Ode=1-\Om-\Ok-\Ogamma$, then interpolated on the test
grid. This inversion enforces the Friedmann constraint on every GP draw, so
the reconstructed $w(z)$ cannot wander into backgrounds that are inconsistent
with a late-time FLRW expansion \cite{Maqsood:2026gpd}. Comoving distances on an extended grid $z\in[0,2000]$ are accumulated by
trapezoidal integration. The recombination sound horizon $r_s(z_*)$, which
depends only on $\Om$ and $\omega_b h^2$ at fixed early-Universe physics, is
evaluated once per MCMC step \cite{Hu:1995en,Eisenstein:1997ik}. The
drag-epoch $r_d$ used for BAO ratios follows a standard fitting formula
consistent with the DESI distance definitions.

\subsection{Sampling and reconstruction}

Flat priors are listed in Table~\ref{tab:priors}. Sampling uses the
\texttt{emcee} ensemble sampler \cite{ForemanMackey:2013,Goodman:2010}. The
baseline chain employs 32 walkers and $10\,000$ steps after a burn-in of
$2000$ steps. Ablation runs use 24 walkers and $4000$ steps (burn-in $1000$),
which is sufficient to compare frameworks even though the effective sample
sizes are smaller (Sec.~\ref{sec:results}). Integrated autocorrelation times
$\tau$ and effective sample sizes $N_{\rm eff}$ are quoted with the results.

\begin{table}[htbp]
\centering
\begin{tabular}{lc}
\toprule
Parameter & Prior range \\
\midrule
$\Hzero$ [km s$^{-1}$ Mpc$^{-1}$] & $[55,80]$ \\
$\Om$ & $[0.05,0.60]$ \\
$\Ok$ & $[-0.05,0.05]$ \\
$\omega_b h^2$ & $[0.018,0.026]$ \\
$\sigma_f$ (zero-mean GP) [\kmsmpc] & $[10,500]$ (sampled in $\ln\sigma_f$) \\
$\sigma_f$ (residual GP) [\kmsmpc] & $[0.1,500]$ (sampled in $\ln\sigma_f$) \\
$\ln l$ (redshift units) & $[\ln 0.3,\ln 5.0]$ \\
\bottomrule
\end{tabular}
\caption{Flat prior ranges. Baseline (zero-mean GP on $H$) and residual-GP
chains share all rows except the amplitude floor.}
\label{tab:priors}
\end{table}
The residual floor $\sigma_f=0.1\,\kmsmpc$ is chosen so that nested $\lcdm$
sits at the lower edge of Eq.~\eqref{eq:residual}; it is not a Bayes factor
for $\Lambda$. All other chains keep $\sigma_f\ge 10$.

Given a posterior sample $\theta^{(j)}$, one GP draw $H^{(j)}(z_*)$ is taken
at $(\sigma_f^{(j)},l^{(j)})$, $E^{(j)}=H^{(j)}/\Hzero^{(j)}$ is formed, and
$f'_{\rm DE}$ is obtained by centred finite differences, so
\begin{equation}
w^{(j)}(z) = -1 + \frac{1+z}{3}\,\frac{f'_{\rm DE}(z)}{f_{\rm DE}(z)}.
\label{eq:wz}
\end{equation}
Medians and 68\% (95\%) credible bands are the corresponding percentiles of the
ensemble $\{w^{(j)}(z)\}$. The same draws yield $f_{\rm DE}(z)$. Low-redshift
summaries quote $w(0)$ as the median over $z<0.15$; the high-redshift error
budget uses the mean 68\% half-width of $w(z)$ for $z>1$.

% =============================================================================
\section{Results}
\label{sec:results}
% =============================================================================

\subsection{Parameter constraints}

For the baseline chain the integrated autocorrelation time reaches
$\tau\simeq 896$ steps, corresponding to about $285$ effective samples per
parameter after burn-in. That $N_{\rm eff}$ is adequate for medians and
68\% intervals; it is tight for quoting a digit-level tension with $w=-1$.
Marginal 16/50/84 percentiles are reported in
Table~\ref{tab:constraints}.

\begin{table*}[htbp]
\centering
\begin{tabular}{lccccc}
\toprule
$\Hzero$ [\kmsmpc] & $\Om$ & $\Ok$ & $\omega_b h^2$ & $\sigma_f$ [\kmsmpc] & $l$ \\
\midrule
$70.1^{+3.6}_{-3.1}$ & $0.317^{+0.032}_{-0.029}$ & $-0.0070^{+0.0013}_{-0.0014}$ & $0.0224^{+0.0001}_{-0.0002}$ & $284.6^{+135.9}_{-72.2}$ & $3.79^{+0.81}_{-1.12}$ \\
\bottomrule
\end{tabular}
\caption{Marginal posterior constraints (68\% C.L.) from the hierarchical
GP analysis of CC+BAO+CMB+SN.}
\label{tab:constraints}
\end{table*}

\begin{figure}[htbp]
\includegraphics[width=\columnwidth]{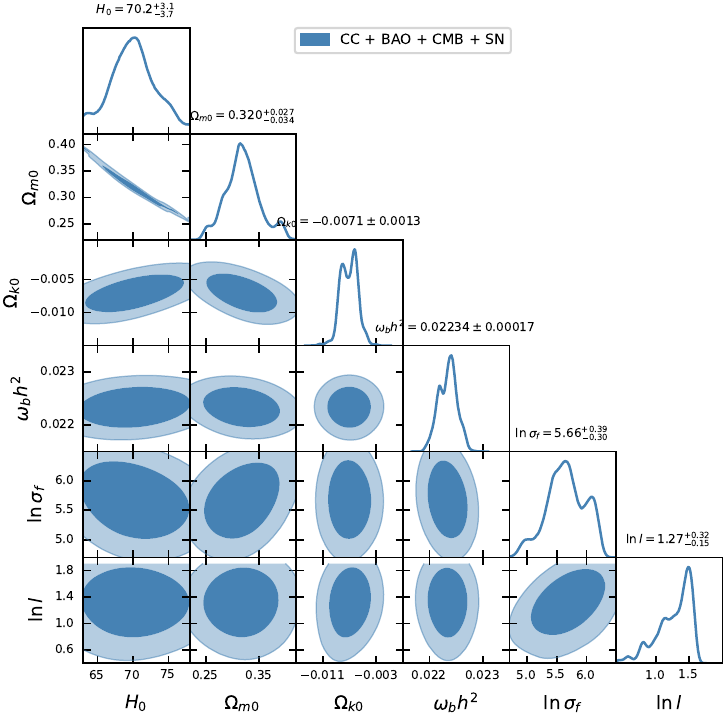}
\caption{Joint posterior of the six parameters (68\% and 95\% contours).
Diagonal panels: 1-D marginals with median and 16th/84th percentiles.
Text and Table~\ref{tab:constraints} quote sample percentiles; small
differences from the getdist titles are from KDE smoothing.}
\label{fig:corner}
\end{figure}

$\Hzero=70.1^{+3.6}_{-3.1}$ lies between Planck
($67.4\pm 0.5$\,\kmsmpc \cite{Planck:2020}) and SH0ES
($73.04\pm 1.04$\,\kmsmpc \cite{Riess:2022}), as in other DESI+CMB
reconstructions
\cite{Loubser:2025fxr,deSouza:2025rhi,Li:2025owk,Wang:2025bkk,Zhang:2025dwu}.
The matter density $\Om=0.317^{+0.032}_{-0.029}$ is consistent with Planck
within the broader GP errors. The curvature constraint
$\Ok=-0.0070^{+0.0013}_{-0.0014}$ is tighter than a full CMB likelihood
typically yields and should not be read as a detection of spatial curvature:
compressed $(R,\ell_A)$ priors together with a GP-reconstructed $D_M(z_*)$
can pin $\Ok$ artificially. We keep $|\Ok|\le 0.05$ as a prior bound and treat
the quoted $\Ok$ error as a pipeline diagnostic. The length scale
$l=3.79^{+0.81}_{-1.12}$ is large compared with the chronometer ML value
(Sec.~\ref{sec:framework}), favouring a smooth $H(z)$; the amplitude is
$\sigma_f=284.6^{+135.9}_{-72.2}$. Figure~\ref{fig:corner} shows limited
degeneracy between $(\ln\sigma_f,\ln l)$ and the cosmological parameters.

\subsection{Reconstructed $w(z)$ and $f_{\rm DE}$}
\label{sec:fde}

\begin{figure}[htbp]
\includegraphics[width=\columnwidth]{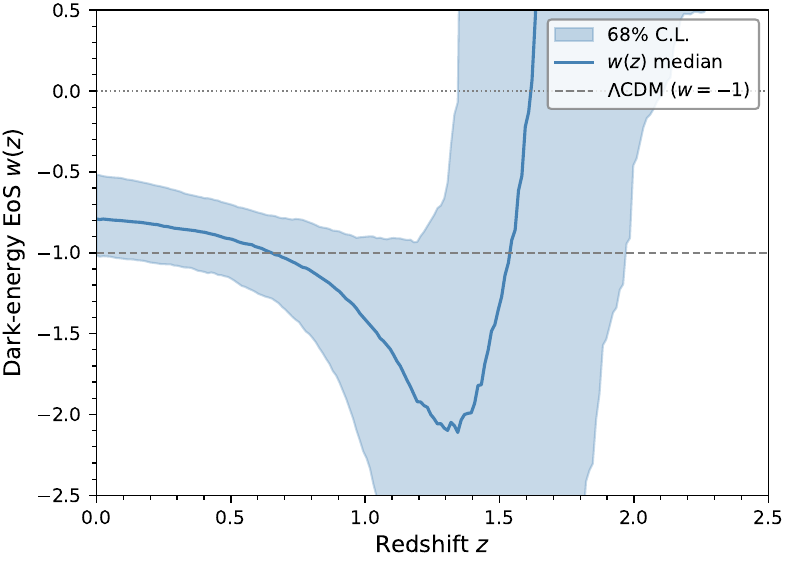}
\caption{Reconstructed dark-energy EoS $w(z)$ from the hierarchical GP
framework. The blue band shows the 68\% credible region; the solid blue
line is the median. The horizontal dashed line marks $\lcdm$ ($w=-1$).}
\label{fig:wz}
\end{figure}

Figure~\ref{fig:wz} shows $w(z)$ with 68\% credible bands. At low redshift,
\begin{equation}
w(z\simeq 0) = -0.80^{+0.26}_{-0.23},
\end{equation}
less than $1\sigma$ from $w=-1$ (the 68\% half-width figure of merit is $0.81$;
with $N_{\rm eff}\sim 285$ we do not treat that digit as a calibrated tension). Strong phantom behaviour ($w\ll -1$) at
$z<0.5$ is excluded at greater than $95\%$ confidence. The band widens at
higher redshift where chronometer leverage thins and BAO+SN provide only
integral constraints; $w=-1$ remains inside the 68\% interval over most of
$0<z<2.5$ (specifically, at every grid point with $z\le 1.6$, covering
$78.5\%$ of the full $z\in[0,2.5]$ grid).

No statistically significant $w(z)\neq -1$ detection appears under this
pipeline. Three elements point in that direction: the large posterior $l$,
which penalises rapid low-$z$ transitions; the CMB+SN distance constraints,
which pull the background toward $\lcdm$-like distances; and the mild
same-data CPL $\Delta\chi^2$ obtained with compressed Planck priors
(Sec.~\ref{sec:framework}).

\begin{figure}[htbp]
\includegraphics[width=\columnwidth]{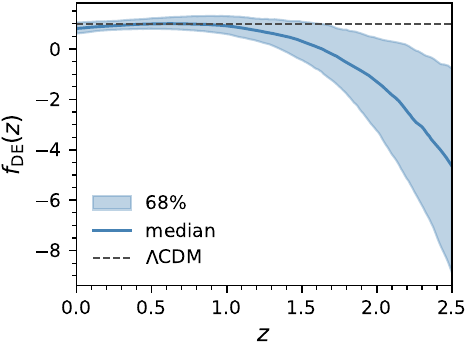}
\caption{Reconstructed normalised dark-energy density $f_{\rm DE}(z)$ from the
hierarchical GP posterior.  The blue band is the 68\% credible region; the
dashed line marks $\lcdm$ ($f_{\rm DE}=1$).}
\label{fig:fde}
\end{figure}

Where $\rho_{\rm DE}$ is weakly constrained, the ratio in Eq.~\eqref{eq:wz2}
amplifies noise, so the normalised density
$f_{\rm DE}(z)=\rho_{\rm DE}(z)/\rho_{\rm DE}(0)$ is the more stable summary
(Fig.~\ref{fig:fde}), in the same spirit as density-space GP assessments of
DESI DR2 \cite{deSouza:2025rhi}. Unity lies inside the 68\% band at
\emph{every} reconstruction node with $z\le 1.6$ (and at $64.5\%$ of the full
grid out to $z=2.5$). The 68\% band first excludes $f_{\rm DE}=1$ at
$z\simeq 1.62$, where chronometer leverage is already thin; the largest
$\lvert\mathrm{median}-1\rvert$ relative to the 68\% half-width on the grid
is $1.4$ at $z\simeq 2.37$ and should not be read as a detection.
At $z=0$, $f_{\rm DE}=0.81^{+0.24}_{-0.20}$, still including unity.
Among the ablations, $f_{\rm DE}=1$ remains inside the 68\% band over the
whole $z\le 1.6$ interval except Union3 and DES-Dovekie, which first leave
unity at $z\simeq 1.42$ and $z\simeq 1.41$ (still covering all $z\le 1$),
and the short-$l$ prior, which first exits at $z\simeq 0.28$. Those early
exits are a kernel or SN-calibration response at the edge of the
chronometer lever arm, not a low-$z$ detection.

\subsection{Framework sensitivity}
\label{sec:framework}

\paragraph{Same-data CPL baseline.}
To separate the role of the GP from that of the data combination, we fit CPL
$(w_0,w_a)$ with free $(\Hzero,\Om,\omega_b h^2)$ on the same DESI DR2 rows
and Planck distance priors, swapping the SN sample, and compare to nested
$\lcdm$ at $(w_0,w_a)=(-1,0)$. Akaike's criterion is
$\mathrm{AIC}=\chi^2+2k$ with $k_\Lambda=3$ and $k_{\mathrm{CPL}}=5$; it does
not involve the data-vector length. BIC uses
$N=n_{\mathrm{BAO}}+n_{\mathrm{CMB}}+n_{\mathrm{SN}}$ as the length of the
Gaussian data vector and is only indicative for correlated supernovae.
We quote $\sqrt{\Delta\chi^2}$ as a one-parameter figure of merit; for two
extra degrees of freedom the nested $p$-values remain $\gtrsim 0.14$ even
for Union3 and DES-Y5.
Table~\ref{tab:sncpl} summarises the nested comparison.
BAO+CMB alone gives $\Delta\chi^2(\lcdm-\mathrm{CPL})\simeq 0.85$
($\Delta\mathrm{AIC}\simeq +3.2$). Pantheon+ leaves the preference mild
($\Delta\chi^2\simeq 0.98$, $\Delta\mathrm{AIC}\simeq +3.0$). Union3 and
DES-Y5 raise $\Delta\chi^2$ to $3.3$ and $4.0$
($\sqrt{\Delta\chi^2}\simeq 1.8$--$2.0$; two-parameter $p\simeq 0.20$ and
$0.14$), while DES-Dovekie brings it back to $1.6$
($\Delta\mathrm{AIC}\simeq +2.4$). In every compressed-CMB combination AIC
still prefers nested $\lcdm$ or is inconclusive (DES-Y5:
$\Delta\mathrm{AIC}\simeq 0.03$). This SN dependence matches independent GP
reconstructions on overlapping stacks
\cite{Maqsood:2026thermo,Dinda:2025jcap,Dinda:2026upd,Popovic:2025dovekie}.
The GP and CPL runs share the compressed pipeline; the mild Pantheon+
$\Delta\chi^2$ is not a statement about DESI's full-likelihood result
\cite{Cortes:2025joz,Wang:2025vfb,Ye:2025ark}. The baseline GP $w(0)$ offset
(less than $1\sigma$) is consistent with the Pantheon+ parametric row. The
hierarchical Union3 swap gives $w(0)=-0.76^{+0.22}_{-0.22}$
($1.11\sigma$ from $-1$), milder than the parametric
$\sqrt{\Delta\chi^2}\simeq 1.8$ on the same sample; DES-Y5 gives
$w(0)=-0.85^{+0.25}_{-0.26}$ ($0.58\sigma$), versus parametric
$\sqrt{\Delta\chi^2}\simeq 2.0$; DES-Dovekie gives
$w(0)=-0.79^{+0.24}_{-0.26}$ ($0.86\sigma$), close to the Pantheon+ GP
(Table~\ref{tab:robust}).

\begin{table}[htbp]
\centering
\begin{tabular}{lcccc}
\toprule
SN sample & $N_\mathrm{SN}$ & $\Delta\chi^2$ & $\Delta\mathrm{AIC}$ & $\sqrt{\Delta\chi^2}$ \\
\midrule
none & 0 & $0.85$ & $+3.15$ & $0.92$ \\
Pantheon+ & 1624 & $0.98$ & $+3.02$ & $0.99$ \\
Union3 (bins) & 22 & $3.25$ & $+0.75$ & $1.80$ \\
DES-Y5 & 1829 & $3.97$ & $+0.03$ & $1.99$ \\
DES-Dovekie & 1820 & $1.59$ & $+2.41$ & $1.26$ \\
\bottomrule
\end{tabular}
\caption{Nested $\lcdm$ versus CPL on the compressed BAO+CMB pipeline,
varying only the supernova sample. $\Delta\chi^2\equiv\chi^2_\Lambda-\chi^2_{\mathrm{CPL}}$;
$\Delta\mathrm{AIC}\equiv\mathrm{AIC}_{\mathrm{CPL}}-\mathrm{AIC}_\Lambda$
(positive prefers $\lcdm$). $N_\mathrm{SN}$ is the data-vector length.
$\sqrt{\Delta\chi^2}$ is a one-parameter figure of merit, not a two-parameter
significance.}
\label{tab:sncpl}
\end{table}

\paragraph{Fixed versus co-sampled hyperparameters.}
Repeating the MCMC with $(\sigma_f,l)$ fixed at the chronometer ML values
($\sigma_f^{\rm ML}\simeq 138$\,\kmsmpc, $l^{\rm ML}\simeq 2.1$) moves
$w(0)$ from $-0.90^{+0.30}_{-0.23}$ (fixed) to the baseline
$-0.80^{+0.26}_{-0.23}$ and changes the mean 68\% half-width of $w(z)$ at
$z>1$ by about $7\%$ (from $3.06$ to $2.85$; Table~\ref{tab:robust}). The
joint posterior prefers $l\simeq 3.8>l^{\rm ML}$, so co-sampling is required for
a fair error budget even when it does not widen every redshift bin
\cite{Crittenden:2011aa}.

\paragraph{Probe ablation and LRG1/2.}
Removing Pantheon+, or dropping the LRG1/2 BAO bins, yields the $w(0)$ rows in
Table~\ref{tab:robust}. Both variants remain within $\lesssim 0.4\sigma$ of
$w=-1$. The $\lcdm$-like conclusion is therefore not driven solely by SN or by
those two LRG measurements in this compressed pipeline.

\paragraph{Kernel and length-scale prior.}
On the chronometer layer alone, the Mat\'ern differentiability sequence
$3/2\to5/2\to7/2\to9/2\to\mathrm{RBF}$ yields a monotonic drop in the
reconstructed $H(z=0)$ from $68.8^{+5.8}_{-6.1}$ to $66.9^{+4.5}_{-4.5}$
(Table~\ref{tab:kernelH0}), with shrinking uncertainties, as expected for an
increasingly smooth covariance prior \cite{Maqsood:2026ker}. Hierarchical
Mat\'ern-$3/2$ is the roughest kernel in that sequence and returns
$w(0)=-1.02^{+0.54}_{-0.50}$ ($0.05\sigma$ from $-1$) with the widest
high-$z$ band ($\langle\Delta w\rangle_{z>1}=3.53$). Mat\'ern-$5/2$ gives
$-1.00^{+0.29}_{-0.28}$ ($0.01\sigma$); Mat\'ern-$7/2$ and $9/2$ give
$-0.87^{+0.28}_{-0.23}$ ($0.52\sigma$) and $-0.88^{+0.29}_{-0.25}$
($0.45\sigma$).
CC-only reconstructions show a monotonic drop in $H(z=0)$ with smoothness
(Table~\ref{tab:kernelH0}). Hierarchical posterior $\Hzero$ does not follow
that sequence one-to-one (Mat\'ern-$3/2$ gives $\Hzero\simeq 65.7$; the RBF
baseline is $70.1$), because BAO+CMB+SN pull $\theta$ jointly with the kernel.
We therefore treat Table~\ref{tab:kernelH0} as a kernel-prior diagnostic on
the chronometer layer, not as a prediction of the hierarchical $\Hzero$.
That shift is a smoothness-prior response, not evidence that one kernel is
preferred by the physics. Restricting $l\le 1.5$ forces more
structure and yields $w(0)\simeq -1.24$, but only at $0.7\sigma$ from $-1$,
with a broader high-$z$ band (Table~\ref{tab:robust}). A short-$l$ prior alone
does not restore a strong quintessence-to-phantom transition of the kind
reported in some DESI CPL analyses
\cite{Keeley:2025stf}. We do not invert a scalar-field potential $V(\phi)$ from
the reconstructed $H(z)$; such post-processing is already common on overlapping
data \cite{Maqsood:2026gpd} and is dominated by derivative noise.

\begin{table}[htbp]
\centering
\begin{tabular}{lcc}
\toprule
Kernel & $H(z=0)$ [\kmsmpc] & $\ln p(\mathrm{CC})$ \\
\midrule
Mat\'ern-$3/2$ & $68.82^{+5.81}_{-6.08}$ & $-157.89$ \\
Mat\'ern-$5/2$ & $68.67^{+5.18}_{-5.28}$ & $-157.90$ \\
Mat\'ern-$7/2$ & $68.33^{+5.09}_{-5.10}$ & $-157.97$ \\
Mat\'ern-$9/2$ & $68.16^{+4.86}_{-4.73}$ & $-158.02$ \\
RBF & $66.92^{+4.51}_{-4.47}$ & $-158.15$ \\
\bottomrule
\end{tabular}
\caption{Chronometer-conditioned GP reconstruction of $H(z=0)$ along the
Mat\'ern differentiability hierarchy. Smoother kernels return lower $H_0$
and narrower errors. Hierarchical MCMC variants for $\nu=3/2,7/2,9/2$ use
the same ablation settings as Mat\'ern-$5/2$ (Sec.~\ref{sec:framework}).}
\label{tab:kernelH0}
\end{table}

\paragraph{Residual GP (nested $\lcdm$).}
Conditioning the GP on $H-H_\Lambda(\theta)$ with
$\sigma_f\in[0.1,500]\,\kmsmpc$ makes $\lcdm$ the $\sigma_f\to 0$ boundary.
An ablation-length residual chain ($24\times 4000$, burn 1000, $N_s=20$)
gives $w(0)=-0.98^{+0.09}_{-0.10}$ ($0.20\sigma$ from $-1$), with
$f_{\rm DE}=1$ inside the 68\% band at every grid point out to $z=2.5$.
The residual amplitude is $\sigma_f=17.9^{+14.1}_{-10.4}\,\kmsmpc$; about
$23\%$ of the posterior lies below the zero-mean-GP prior floor
$\sigma_f=10$. The same chain prefers a high-$\Om$, low-$\Hzero$ background
($\Hzero=61.5^{+6.5}_{-4.4}$, $\Om=0.41^{+0.07}_{-0.08}$) that sits near the
lower $\Hzero$ prior edge. That is a compressed-distance plus nested-mean
degeneracy, not a late-time dark-energy detection, and the $\sigma_f$
marginal is not a Bayes factor between $\lcdm$ and a free $H(z)$. Nested
$\lcdm$ remains a good description of the same compressed data once the GP
is not forced to model the entire $H(z)$.

\begin{table}[htbp]
\centering
\begin{tabular}{lccc}
\toprule
Variant & $w(0)$ / $(w_0,w_a)$ & $\sigma_{-1}$/$N\sigma$ & $\langle\Delta w\rangle_{z>1}$ \\
\midrule
Hierarchical (co-sampled) & $-0.80^{+0.26}_{-0.23}$ & $0.81$ & $2.85$ \\
Fixed hypers (CC ML) & $-0.90^{+0.30}_{-0.23}$ & $0.36$ & $3.06$ \\
No SN (CC+BAO+CMB) & $-0.91^{+0.28}_{-0.28}$ & $0.33$ & $2.90$ \\
No LRG1/2 BAO & $-0.89^{+0.31}_{-0.27}$ & $0.36$ & $2.95$ \\
Union3 SN & $-0.76^{+0.22}_{-0.22}$ & $1.11$ & $2.94$ \\
DES-Y5 SN & $-0.85^{+0.25}_{-0.26}$ & $0.58$ & $2.93$ \\
DES-Dovekie SN & $-0.79^{+0.24}_{-0.26}$ & $0.86$ & $2.96$ \\
Mat\'ern-$3/2$ kernel & $-1.02^{+0.54}_{-0.50}$ & $0.05$ & $3.53$ \\
Mat\'ern-$5/2$ kernel & $-1.00^{+0.29}_{-0.28}$ & $0.01$ & $3.16$ \\
Mat\'ern-$7/2$ kernel & $-0.87^{+0.28}_{-0.23}$ & $0.52$ & $2.99$ \\
Mat\'ern-$9/2$ kernel & $-0.88^{+0.29}_{-0.25}$ & $0.45$ & $3.02$ \\
Short-$l$ prior ($l\le 1.5$) & $-1.24^{+0.37}_{-0.36}$ & $0.66$ & $3.24$ \\
Residual GP ($H-H_\Lambda$) & $-0.98^{+0.09}_{-0.10}$ & $0.20$ & $1.16$ \\
CPL BAO+CMB & $(-0.96,-0.28)$ & $0.92$ & --- \\
CPL BAO+CMB+SN & $(-0.94,-0.32)$ & $0.99$ & --- \\
\bottomrule
\end{tabular}
\caption{Framework-sensitivity summary from the hierarchical pipeline and a
same-data CPL fit.  $w(0)$ uses $z<0.15$; $\sigma_{-1}$ is the deviation
from $w=-1$ in units of the 68\% half-width.  For CPL rows, $N\sigma$ is the
one-parameter figure of merit $\sqrt{\Delta\chi^2}$ versus nested $\lcdm$.
$\langle\Delta w\rangle_{z>1}$ is the mean 68\% half-width of $w(z)$ for
$z>1$.  Ablation rows use $24\times 4000$ MCMCs (burn 1000), shorter than the
production baseline ($32\times 10000$) but adequate for framework comparison;
the Mat\'ern and short-$l$ chains have effective sample sizes
$N_{\rm eff}\sim 150$ and $\sim 120$, respectively; SN-swap and additional
kernel rows use the same sampler settings and are intended for the same
framework comparison.}
\label{tab:robust}
\end{table}

% =============================================================================
\section{Discussion}
\label{sec:disc}
% =============================================================================

Under compressed Planck distance priors, CPL remains $\lesssim 1\sigma$ from
$\lcdm$ for Pantheon+ and DES-Dovekie. Union3 and DES-Y5 raise
$\sqrt{\Delta\chi^2}$ to $\simeq 1.8$--$2.0$, but the two-parameter nested
$p$-values stay $\gtrsim 0.14$, and AIC still does not prefer CPL
(Table~\ref{tab:sncpl}). The
hierarchical GP is less than $1\sigma$ from $w=-1$ on Pantheon+, $1.1\sigma$ on
Union3, $0.6\sigma$ on DES-Y5, and $0.9\sigma$ on DES-Dovekie.
Those statements concern this pipeline, not DESI's CPL fit with the full Planck
likelihood \cite{DESI:2025}. The contrast with that headline significance
tracks CMB compression and related analysis choices
\cite{Cortes:2025joz,Wang:2025vfb,Ye:2025ark}.
GP reconstructions that place $w(z)$ inside a Boltzmann solver with the full
CMB likelihood find stronger deviations \cite{Lodha:2025qbg} than
compressed-CMB GPs on chronometers+BAO+SN
\cite{Maqsood:2026gpd,deSouza:2025rhi}. Table~\ref{tab:robust} isolates
SN, LRG1/2, kernel, and $l$-prior switches inside a single pipeline so that
those modelling differences can be compared on equal footing.

The reconstructed $w(z\simeq 0)=-0.80^{+0.26}_{-0.23}$ is compatible with the
Pantheon+-only GP of Ref.~\cite{Benisty:2023}
($w(0)=-1.01^{+0.15}_{-0.14}$) within uncertainties, and with recent
compressed-CMB GP reconstructions that remain consistent with $w=-1$
\cite{Maqsood:2026gpd}. The DESI CPL point
$(w_0,w_a)\approx(-0.8,-0.7)$ \cite{DESI:2025} sits near the edge of our
low-$z$ 68\% band, but would require stronger high-$z$ evolution than either
the GP or the same-pipeline CPL supports. Density space makes the same point
more cleanly: $f_{\rm DE}=1$ is inside the 68\% band throughout $z\le 1.6$,
in line with the 95\% model-compatibility conclusion of
Ref.~\cite{deSouza:2025rhi}. In that sense the hierarchical
reconstruction and the compressed CPL fit tell a consistent story: the data
combination used here does not demand a large $w_a$.

Co-sampling $(\sigma_f,l)$ follows the hierarchical philosophy of
\cite{Crittenden:2011aa}. Earlier GP cosmography often fixed hyperparameters
after an ML step
\cite{Seikel:2012,Holanda:2017,Busti:2014dua,Niu:2023bsr}. The
fixed-hyperparameter row of Table~\ref{tab:robust} shows that the high-$z$
error budget changes at the several-percent level, while the low-$z$ median
moves by $\sim 0.1$ in $w$. Ablation MCMCs are shorter
($24\times 4000$, burn 1000) than the production baseline
($32\times 10\,000$), with $N_{\rm eff}\sim 150$ (Mat\'ern-$5/2$) and
$\sim 120$ (short-$l$); later kernel and SN-swap rows use the same ablation
length and are intended for framework comparison rather than
for percent-level posterior calibration.

Several limitations should be kept in view.
Compressed CMB priors discard information that the full Planck likelihood
retains; a future run with the latter would be needed before claiming tension
or concordance with DESI CPL analyses that use the full likelihood.
The GP is conditioned only on chronometer $H(z)$ and is frozen above $z=2.5$
when integrating to recombination, so early-time physics enters solely through
$r_s$ and the distance priors: high-$z$ $w(z)$ is not a CMB reconstruction.
The Monte-Carlo $\mathcal{L}_{\eff}$ with finite $N_s$ adds sampling noise to
each likelihood call; we have not separately quantified that noise on $w(0)$.
The analysis is background-only: without growth data we cannot distinguish a
time-varying $w(z)$ from a dark-sector interaction that mimics the same $H(z)$
\cite{You:2026dsd}.
SN calibration is first-order: Union3 and DES-Y5 raise the compressed
CPL $\Delta\chi^2$ relative to Pantheon+, while DES-Dovekie reduces it again
(Table~\ref{tab:sncpl}), consistent with calibration-independent BAO--SN tests
\cite{Dinda:2025jcap,Dinda:2026upd,Popovic:2025dovekie}. The hierarchical GP
tracks that ordering but stays milder ($1.1\sigma$, $0.6\sigma$, $0.9\sigma$).
The production model is a zero-mean GP on $H$ [Eq.~\eqref{eq:zeromean}], so
part of the offset from $\Lambda$ is prior-supported; the residual nest
[Eq.~\eqref{eq:residual}] is a diagnostic, not a Bayes factor, and its
$\Hzero$ should not be read as a Hubble-tension result.
Natural extensions include the full DESI BAO covariance updates and a joint
analysis with the complete Planck likelihood.

% =============================================================================
\section{Conclusions}
\label{sec:conc}
% =============================================================================

We reconstruct late-time expansion with a hierarchical GP that co-samples
$(\sigma_f,l)$ and couples CC, DESI DR2 BAO, compressed Planck distance
priors, and Pantheon+ through a Monte-Carlo effective likelihood. The
baseline result is $w(z\simeq 0)=-0.80^{+0.26}_{-0.23}$
(less than $1\sigma$ from $w=-1$) with $l\simeq 3.8$. The reconstructed
$f_{\rm DE}=1$ remains inside the 68\% band at every node with $z\le 1.6$.
On the same compressed data, CPL improves nested $\lcdm$ by only
$\Delta\chi^2\sim 1$, while AIC prefers nested $\lcdm$. Probe,
tracer, kernel, SN, and length-prior ablations leave $w=-1$ viable at
$\lesssim 1\sigma$. A residual GP nested on $\lcdm$ is $0.2\sigma$ from $w=-1$;
that chain is a diagnostic, not a replacement of the production baseline.
Kernel choice is treated as a smoothness prior
\cite{Maqsood:2026ker}. This stack does not require a large $w_a$. The numbers
are tied to compressed CMB information and to the hierarchical setup; they are
not a test of DESI's full-likelihood CPL preference. The quoted $\Ok$ error
and the residual-GP $\Hzero$ should be read as compressed-prior diagnostics,
not as cosmological measurements.

\acknowledgments
E.L. acknowledges support from the National Natural Science Foundation of
China under Grant No.~12273001.
This work used publicly released data from DESI, Planck, Pantheon+, Union3,
and the Dark Energy Survey.

\paragraph{Data availability.}
The DESI DR2 BAO measurements, Planck 2018 compressed distance priors, Pantheon+,
Union3, DES-Y5 and DES-Dovekie supernova catalogues, and cosmic-chronometer
compilation used in this work are publicly available from the respective
collaborations and references cited in Sec.~\ref{sec:method}.

\bibliographystyle{unsrtnat}
\bibliography{references}

\end{document}